\documentclass[a4paper,11pt,headings=big,DIV=12]{article}
\pdfoutput=1 
\usepackage{amsmath,amssymb,mathtools}

\usepackage[usenames, dvipsnames]{color}
\usepackage{tcolorbox} 
\definecolor{mygray}{gray}{0.2}
\usepackage{bbold}

\usepackage{jheppub}

\usepackage{bm,amssymb,slashed,graphicx,multirow,soul,mathtools,xspace,array}  
\usepackage{float}                   
\allowdisplaybreaks
\usepackage{ bbold }
\usepackage{subfigure}
\usepackage{slashed}
\usepackage{acronym}

\definecolor{violet}{rgb}{0.94, 0.2, 0.8}
\definecolor{lightblue}{rgb}{0.39, 0.58, 0.93} 
\definecolor{asparagus}{rgb}{0.53, 0.66, 0.42}

\acrodef{PDG}[PDG]{Particle Data Group}
\acrodef{OPE}[OPE]{Operator Product Expansion}
\acrodef{FCNC}[FCNC]{flavour-changing neutral current}
\acrodef{RHC}[RHC]{right-handed currents}
\acrodef{SM}[SM]{Standard Model}
\acrodef{NP}[NP]{New Physics}
\acrodef{MFV}[MFV]{Minimal Flavour Violation}
\acrodef{SD}[SD]{short-distance}
\acrodef{LD}[LD]{long-distance}
\acrodef{DA}[DA]{distribution amplitude}

\newcommand{\vev}[1]{\langle #1 \rangle} 
\newcommand{\state}[1]{|#1\rangle}

\newcommand{\matel}[3]{\langle #1|#2|#3\rangle}

\newcommand{\al}{\alpha}
\newcommand{\be}{\beta}
\newcommand{\ga}{\gamma}

\newcommand{\de}{\delta}
\newcommand{\De}{\Delta}
\newcommand{\la}{\lambda}
\newcommand{\La}{\Lambda}

\newcommand{\sig}{ \sigma}

\newcommand{\MeV}{\,\mbox{MeV}}

\newcommand{\pl}{\!+\!}

\newcommand{\EQ}{Eq.~}
\newcommand{\EQs}{Eqs.~}

\newcommand{\FIG}{Fig.~}

\newcommand{\SEC}{Sec.~}
\newcommand{\SECs}{Secs.~}
\newcommand{\APP}{App.~}
\newcommand{\APPs}{Apps.~}

\newcommand{\gast}{\ga_*}
\newcommand{\best}{\be_*}

\newcommand{\CDQCD}{\text{CD-QCD}}

\newcommand{\CFT}{\text{CFT}}

\newcommand{\OSa}{{S^a}}

\newcommand{\OPa}{{P^a}}

\newcommand{\Tr}{\mathrm{Tr}}

\newcommand{\ORD}{{\cal O}}

\newcommand{\dil}{D}

\newcommand{\mq}{m_q}

\newcommand{\Tud}[2]{T^{#1}_{\;\; #2}}
\newcommand{\TEMT}{ \Tud{\rho}{\rho} }
\newcommand{\veci}[1]{\vec{#1}}  

\newcommand{\mink}{\eta}

\newcommand{\Oone}{O_1}
\newcommand{\Otwo}{O_2}

\newcommand{\phiIR}{\phi_{\text{IR}}}
\newcommand{\Nf}{N_f}
\newcommand{\Nc}{N_c}
\newcommand{\Op}{{\cal O}}

\definecolor{violet}{rgb}{0.94, 0.2, 0.8}
\definecolor{lightblue}{rgb}{0.39, 0.58, 0.93} 
\definecolor{lightgreen}{rgb}{0.1, 0.73, 0.33}

\DeclareOldFontCommand{\tt}{\normalfont\ttfamily}{\mathtt}

\begin{document}
\preprint{CERN-TH-2023-100}

\title{\boldmath QCD with an Infrared Fixed Point - Pion Sector }

\author[1,2]{Roman Zwicky,}

\affiliation[1]{Higgs Centre for Theoretical Physics, School of Physics and
Astronomy, \\ The University of Edinburgh, 
Peter Guthrie Tait Road, Edinburgh EH9 3FD, Scotland, UK}
\affiliation[2]{ Theoretical Physics Department, CERN, \\
Esplanade des Particules 1,  Geneva CH-1211, Switzerland}
\emailAdd{Roman.Zwicky@ed.ac.uk}
\abstract{The possibility that gauge theories with chiral symmetry breaking 
below the  conformal window exhibit 
an infrared fixed point  is explored.  
With this assumption three aspects of pion physics are reproduced if 
the the quark mass anomalous dimension at the infrared fixed point  is $\gamma_* = 1$: 
First, by matching the long-distance  scalar adjoint  correlation function. 
Second, by perturbing the fixed point by a small quark mass, the  $m_q$-dependence 
of the pion mass  is reproduced by renormalisation group arguments. Third, 
consistency of the trace anomaly 
and the Feynman-Hellmann theorem, for small $m_q$, imply the same result once more.
 This suggests the following picture for the conformal window: close to its upper boundary 
 $\gamma_*$ is zero 
and grows as the number of fermions is reduced until its lower boundary
$\gamma_*=1$ is reached,  
where chiral symmetry breaking sets in.  
Below, the strongly coupled gauge theory with $\gamma_*=1$ is infrared dual to the free theory of pions. 
A possible dilaton sector of the scenario will be addressed in a  companion paper.
}

\maketitle

\flushbottom

\setcounter{tocdepth}{3}
\setcounter{page}{1}
\pagestyle{plain}

\section{Introduction}

The idea that  spontaneous  chiral symmetry breaking in the strong interaction  
induces scale spontaneous   symmetry breaking (SSB) predates QCD \cite{Isham:1970xz,Isham:1970gz,
Ellis:1970yd,Ellis:1971sa,Crewther:1970gx,Crewther:1971bt}.
The goal of this paper is to explore this idea within gauge theories, 
using parts of chiral perturbation theory ($\chi$PT) \cite{Gasser:1983yg,Donoghue:1992dd,Leutwyler:1993iq,Scherer:2012xha}
and the renormalisation group (RG).  
Whether this scenario corresponds to a new phase \cite{DelDebbio:2021xwu},
or an unexplored feature of QCD  has to be left open at this stage.  
The assumption of an infrared fixed point (IRFP) is  non-standard. The main point 
of the paper is that under this hypothesis  aspects of pion physics 
are reproduced consistently.
 
An   IRFP  and  scale  SSB is accompanied by a (pseudo) Goldstone boson, 
known as the dilaton. 
Its features and interactions are less transparent than that of the pion  as scale symmetry is only emergent in the IR. 
Since the results presented here are seemingly independent of dilaton-aspects, 
its main discussion is postponed to a companion paper \cite{RZdilaton}.\footnote{Possibly the most spectacular aspect of a dilaton 
is that massive hadrons, such as a nucleon,   and a traceless energy momentum tensor (EMT) $\matel{N}{\TEMT}{N}=0$ are compatible with each other.
 \cite{DelDebbio:2021xwu}.  
The dilaton restores the dilatation Ward identity just as the pion does for chiral Ward identities. 
Another attractive feature is that the gauge theory contribution to the cosmological constant could be zero for $\mq = 0$ \cite{DelDebbio:2021xwu}.
It is worthwhile to mention that the dilaton under discussion is not a  gravity-scalar, such as in  string theory, 
nor an accidentally light scalar but a genuine Goldstone resulting from SSB, e.g. \cite{Crewther:2020tgd} for a historical perspective. If QCD were to possess an IRFP and a dilaton, there is consensus that it corresponds to 
the $\sig$-meson, known as the $f_0(500)$ in the Particle Data Group \cite{PDG22}.}  
At the end of the paper, we briefly comment on   how the addition of a dilaton does not alter the results.

The starting assumption is  that the massless degrees of freedoms, to which we will refer to as IR-states, 
see the world as a conformal field theory (CFT) in the deep-IR.\footnote{See  \APP\ref{app:conscale} for comments on scale versus conformal symmetry.}  That is, the trace of the EMT on the IR-states $\phiIR$
\begin{equation}
\label{eq:IRconformality}
\matel{\phiIR(p)}{\TEMT}{\phiIR'(p)} \to 0 \;,
\end{equation}
vanishes for zero momentum transfer.\footnote{This does not imply that any definition of a $\be$-function  assumes a zero in the IR as it is only the combination of  $\be$ times the field strength tensor which is RG-invariant cf. \SEC\ref{sec:RG}. 
This aspect has  for example been emphasised in the review \cite{Nogradi:2016qek}.} 
 It is though reasonable to assume that there exists a scheme for which $\best=0$ if \eqref{eq:IRconformality} holds.  
Amongst those degrees of freedom are the vacuum, the pions resulting from chiral SSB and possibly the dilaton \cite{DelDebbio:2021xwu,RZdilaton}.  \EQ \eqref{eq:IRconformality} may be regarded 
as the minimal form by which IR-conformality manifests itself in the dilatation Ward identity. 
Technically  this means  that  correlation functions  in the deep-IR, and generally physical observables, are determined by the scaling dimension  $\vev{{\cal O}(x){\cal O}(0)} \propto (x^2)^{-\Delta_{\cal O}}$.  The quark mass anomalous dimension 
$\ga_m$, denoted by $\gast = \ga_m|_{\mu=0}$ at the IRFP, 
governs the scaling dimension of many important operators. 
The central result of this paper  is that with the IRFP-assumption, this anomalous 
dimension must assume 
 \begin{equation}
 \label{eq:gast}
\gast = 1 \;.
\end{equation}
This is inferred in three different ways, by matching the pion low energy physics with the gauge theory. 
The value \eqref{eq:gast} is then important in two respects: it marks the lower boundary of the conformal window \emph{and} it describes the pion physics in the chirally broken phase in 
terms of the strongly coupled IRFP of the gauge theory. Whereas the former is compatible with previous work 
and lattice Monte-Carlo studies, as discussed in the conclusions, the latter is a new perspective.
 \begin{centering}
\begin{figure}[h!]
\includegraphics[width=1.0\linewidth]{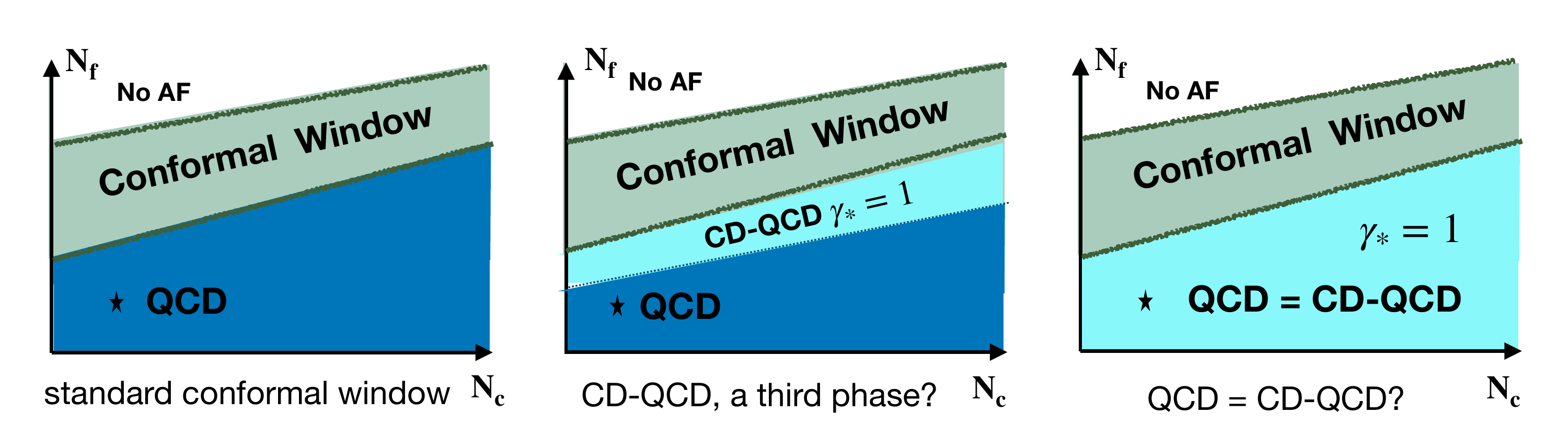}
\vspace{-0.5cm}
	\caption{\small 
	Sketch of phase diagrams of gauge theories with quark matter as  a function of 
	the number of flavours $N_f$ and colours $N_c$, as described in the main text.  
	``No AF" stands for no asymptotic freedom and its boundary is known 
	from the Caswell-Banks-Zaks analysis \cite{Caswell:1974gg,Banks:1981nn}.
	 The lower dark green 
	line marks the end of the conformal window and its precise location is unknown 
	in the non-supersymmetric case.  
	In the lower dark blue phase chiral symmetry is broken, hadrons confine  and $N_f = 3$ and $N_c =3$ 
	represents QCD.  
	 (left) Literature-standard conformal window scenario. (centre)  CD-QCD as a third phase  as advertised
	  in \cite{DelDebbio:2021xwu}.   
	   (right) QCD and CD-QCD are one and the same.  We emphasise again that
	   boundaries other than the one of AF are unknown and are shown for illustrative purposes only.}
	\label{fig:overview}  
\end{figure}
\end{centering}

The main part of the paper  consists of \SEC\ref{sec:RGpion} where  $\gast =1$ \eqref{eq:gast}  
is derived from:  a) a specific long-distance correlator, b)  the hyperscaling relation of the pion mass,  
and c) the matching of the trace anomaly with the Feynman-Hellmann theorem, given in  \SECs  \ref{sec:deep}, \ref{sec:gast} and \ref{sec:FHTEMT}. respectively. In \SEC\ref{sec:dilaton} we comment on what happens when a dilaton is added.
The paper ends with a summary and discussion in \SEC\ref{sec:conc}. 
\APPs\ref{app:conv}, \ref{app:conscale} and \ref{app:soft} contain conventions, 
related discussion of scale versus conformal invariance and the soft-pion theorem in use. 

\section{Consequences of an IRFP for QCD-like Theories}
\label{sec:RGpion}

 The conformal window is reviewed  as this work builds on it,  and for further reading the reader is referred to 
 \cite{Sannino:2009za,DelDebbio:2010zz,Nogradi:2016qek}.
 The starting point is an asymptotically free gauge theory with gauge group $G$, e.g.  
 $G=SU(N_c)$,   and $N_f$ massless quarks in a given representation of $G$.  
The point of study are the IR phases of these gauge theories as a function of  $\Nc$, $\Nf$ and the quark 
representation, cf. \FIG\ref{fig:overview}. 
The figure on the left depicts the standard picture for non-supersymmetric gauge theories.\footnote{\label{foot:SUSY} The conformal window of supersymmetric theories is well-understood  due to the Seiberg dualities 
\cite{Intriligator:1995au,Shifman:1999mk,Terning:2006bq,Ryttov:2017khg} and the exact NSVZ $\be$-function \cite{Novikov:1983uc}.
 The lower boundary becomes 
a perturbative FP in the dual theory.  The value of $\gast$ at which the transition occurs is $\gast =1$, related 
to the unitarity bound of the squark composite operator $\De_{\tilde{Q}Q} = 2 - \gast \geq 1$. 
The phases below the conformal window are richer in that there is a phase with confinement without chiral symmetry breaking. It is not believed that this is repeated for non-supersymmetric gauge theories.}
The boundary in the $(\Nc,\Nf)$-plane of where asymptotic freedom is lost is known and  for  $N_f$ below the boundary 
the theories admit a perturbative IRFP, the so-called Caswell-Banks-Zaks FP  \cite{Caswell:1974gg,Banks:1981nn}. 
This phase, shown in green, continues until the coupling becomes strong enough for chiral symmetry to break via 
the formation of the quark condensate $\vev{\bar qq} \neq 0$, marked in dark blue and collectively referred to as QCD. 
This breaks the global flavour symmetry $SU(\Nf)_L \otimes SU(\Nf)_R \to SU(\Nf)_V$, accompanied by 
$N_f^2-1$ massless pions as Goldstones and is believed to cause 
 quarks and gluons to confine into hadrons.  The exact boundary between the two phases is unknown and the matter of intensive debates in the literature. 
All evidence points towards a monotonically increasing $\gast$, cf. the list of references in the conclusions. 
A large   $\gast$ is  important for the walking technicolor scenario, e.g. \cite{Hill:2002ap,Sannino:2009za}, 
and gave rise to  efforts to determine it from lattice Monte Carlo simulations, e.g. \cite{Hasenfratz:2020ess,Kuti:2022ldb,Fodor:2019ypi,Fodor:2017gtj,Fodor:2018uih,Fodor:2019vmw,Fodor:2020niv,
Chiu:2018edw,LatticeStrongDynamics:2018hun,DelDebbio:2015byq,Brower:2015owo,LSD:2014nmn,LatKMI:2014xoh,LatKMI:2016xxi} as
reviewed in  \cite{DelDebbio:2010zz,Nogradi:2016qek,Witzel:2019jbe}.
In   \cite{DelDebbio:2021xwu}  the conformal dilaton was advocated as a third phase as shown 
in the central figure and its domain and location should not be taken literally.  
In this work we refer to this phase as the conformal dilaton CD of QCD. 
It seemed reasonable to assume 
that this phase lies in between the others as it is the same for its properties.  Clearly neither its existence nor its location 
are certainties.  Any of the three cases shown in  \FIG\ref{fig:overview} are  logical possibilities. 
This paper consists in analysing the IRFP scenario,  or the CD-QCD phase. 
It seems worthwhile to reemphasise  that none of the results obtained directly depend on the presence of a dilaton.

\subsection{Deep-IR interpretation of the adjoint scalar correlator   ($\mq = 0$)}
\label{sec:deep}

For $\mq = 0$ the theory exhibits, the previously mentioned,  
scaling in correlation functions  and this is what we will exploit in this section. 
The  scalar operator, with   $J^{P}  = 0^{+}$ quantum numbers
\begin{equation}
\label{eq:Ops}
 \OSa = \bar q   T^a q \;,
\end{equation}
where $T^a$ generates the  flavour symmetry,
is an example that offers itself since it is not perturbed by a single Goldstone. 
 Consistency of the IRFP interpretation means that
\begin{equation}
\label{eq:matchSa}
 \vev{\OSa(x) \OSa(0) }_{\CDQCD} =   \vev{\OSa(x) \OSa(0) }_{\chi\text{PT}} \;, \quad \text{for } x^2 \to \infty \;,
\end{equation}
must hold, since they describe the same theory in the deep-IR limit $x^2 \to \infty$. 
The next two sections are devoted to this matching.\footnote{This correlator has been 
used in \cite{Smilga:1993in} to match $\chi$PT and the spectral representation. 
It was deduced that the correction to the Dirac eigenvalue density is $\rho(\la) - \rho(0) = 
C |\la|$, where $C$ is known and $\rho(0) = -\vev{\bar qq}/\pi$ is the famous Banks-Casher relation \cite{Banks:1979yr}.}

\subsubsection{The CD-QCD correlator in the deep-IR}

It is our assumption that QCD is described by an IRFP in the deep-IR, 
which in turn means that CFT methods apply in that regime. 
 In CFTs 2- and 3-point  correlators  \cite{Mack:1988nf,DiFrancesco:1997nk,Braun:2003rp,HOCFT,Rychkov:2016iqz,Poland:2018epd} are entirely  governed by their scaling dimensions, $\De_{\Op} = d_{\Op} + \ga_{\Op}$, which is the sum of the engineering dimension and the anomalous dimension.
Concretely, for a Euclidean CFT
\begin{equation}
\label{eq:CFT}
\vev{\Op(x) \Op^\dagger(0) }_{\CFT} \propto (x^2)^{-\De_\Op} \;,
\end{equation}
where $x^2 = x_0^2 + x_1^2+x_2^2 +x_3^2$ 
and $\vev{\dots}$ denoting, hereafter, the vacuum expectation value.  
The behaviour in \eqref{eq:CFT} should be mirrored by the  correlation function \eqref{eq:matchSa} in the deep-IR.
The only necessary ingredient is the 
  scaling  dimension of $\OSa$ which is  
\begin{equation}
\label{eq:Dels}
  \De_{ \OSa}  =   d_{ \OSa} - \gast =  3- \gast  \;, 
\end{equation}
since $d_{ \OSa} =3$.   
\EQ \eqref{eq:Dels} follows from  $\De_{ \OSa}= \De_{ \OPa}$, which holds at least in perturbation theory 
since the $\ga_5$ can be commuted through the diagram for $ \OPa  = \bar q i \ga_5  T^a q $ 
to recover $\OSa$ if $\mq =0$ is assumed. In turn, $\De_{ \OPa} = 3- \ga_m$ follows from the Ward identity 
$\partial^\mu \vev{A^a_\mu(x) P^b(0)} \propto \de^{(4)}(x)  \de^{ab} \vev{\bar qq}$ and the fact that 
$A^a_\mu$ and  $\mq \bar qq$ and  are RG invariants. This is true for the former 
since it is a softly conserved current and for the latter 
it follows for instance from the quantum action principle  for 
which the reader is referred to \cite{Collins:1976yq}, for a discussion in the perturbative context.  
With \eqref{eq:CFT} and \eqref{eq:Dels}, one concludes that 
\begin{equation}
\label{eq:S-QCD}
\vev{\OSa(x) \OSa(0) }_{\CDQCD} \propto  (x^2)^{- (3-\gast)  }   \;, \quad x^2 \to \infty  \;.
\end{equation}

 \begin{figure}[t]
 \begin{center}
\includegraphics[width=0.8\linewidth]{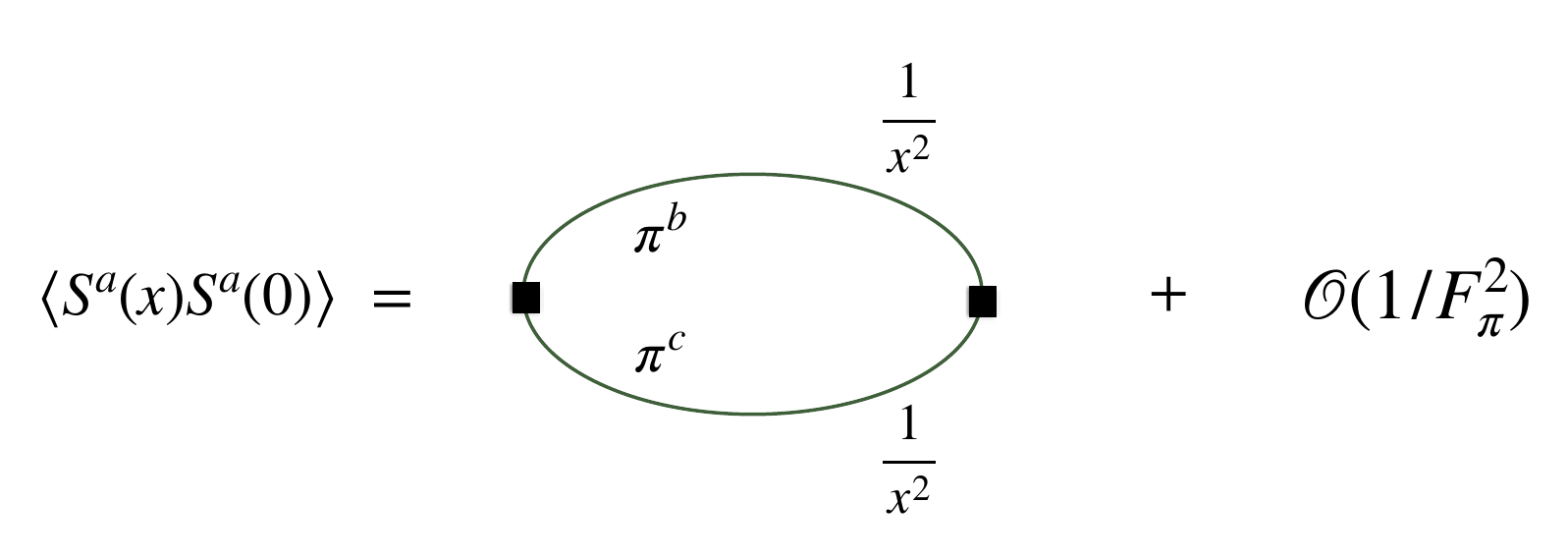}
\vspace{-0.5cm}
	\caption{\small  Adjoint scalar 
	correlation function in $\chi$PT \eqref{eq:chiPT} which  behaves as $1/x^4$ for large distances. }
	\label{fig:dias}
	\end{center}
\end{figure}

\subsubsection{Leading order chiral perturbation theory}

In order to compute the correlator \eqref{eq:matchSa} in $\chi$PT,  the QCD-operator $\OSa$ needs to be described 
 in terms of pion fields.  This can be done by the source method  \cite{Gasser:1983yg,Donoghue:1992dd,Leutwyler:1993iq,Scherer:2012xha}, starting from the LO mass Lagrangian 
\begin{equation}
\de {\cal L}_{\mq} = \frac{F_\pi^2 B_0}{2} \Tr[ {\cal M} U^\dagger + U {\cal M}^\dagger] \;,
\end{equation}
where $B_0 = -\vev{\bar qq}/F_\pi^2$ and the quark mass matrix is ${\cal M} = \mq \mathbb{1}_{\Nf}$ in our case. 
The operator $S^a$ is obtained by replacing ${\cal M} \to T^a J_{S^a}$ and differentiating the log of the Euclidean generating functional ${\cal Z}$ 
 w.r.t. to the source, $\vev{S^a(x)} \leftrightarrow \de_{J_{S^a}(x)} \ln {\cal Z}$,  
\begin{equation}
\label{eq:JSa}
S^a = - \frac{F_\pi^2 B_0}{2} \Tr[T^a U^\dagger + U T^a] \propto B_0 d^{abc} \pi^b \pi^c +  \ORD(1/F_\pi^2) \;,
\end{equation}
where the $ \ORD(1/F_\pi^2)$-terms are cutoff-suppressed and thus  next-LO.\footnote{ 
Generically, $d^{abc} \neq 0$ but for $N_f =2$ it vanishes;
$d^{abc} d^{abc} \propto \Nf^2-4$. 
This accidentality is of 
no special concern to the argument made in this section.}  
The computation of the  correlator in LO $\chi$PT is now straightforward
\begin{equation}
\label{eq:chiPT}
\vev{\OSa(x) \OSa(0) }_{\chi\text{PT}}  \propto B_0^2 d^{abc} d^{abc}  \vev{\pi^e(x) \pi^e(0)}^2  \propto \frac{1}{x^4}  
\;, \quad \text{for } x^2 \to \infty \;,
\end{equation}
where as anticipated $\mq \to 0$ limit has been assumed.
Above $  \vev{\pi^e(x) \pi^e(0)} = 
   \frac{1}{(4\pi)^2}\frac{1}{x^2}$, with $e$ fixed, is  the standard  Euclidean  propagator for a massless scalar field $\pi^e(x)$. 
  Thus the LO $\chi$PT is just given by a free field theory computation as illustrated in \FIG\ref{fig:dias}.

 \subsubsection{IRFP-matching and  contemplation on $\gast=1$ in the wider picture}
 \label{sec:match}
 
 The matching of CD-QCD and $\chi$PT, as in \eqref{eq:matchSa} ,  with 
 \eqref{eq:S-QCD} and \eqref{eq:chiPT} enforces 
  \begin{equation}
 \label{eq:main}
 \gast = 1 \;,
 \end{equation}
 which is the main result of this work.
 
\begin{figure}[t]
 \begin{center}
\includegraphics[width=0.8\linewidth]{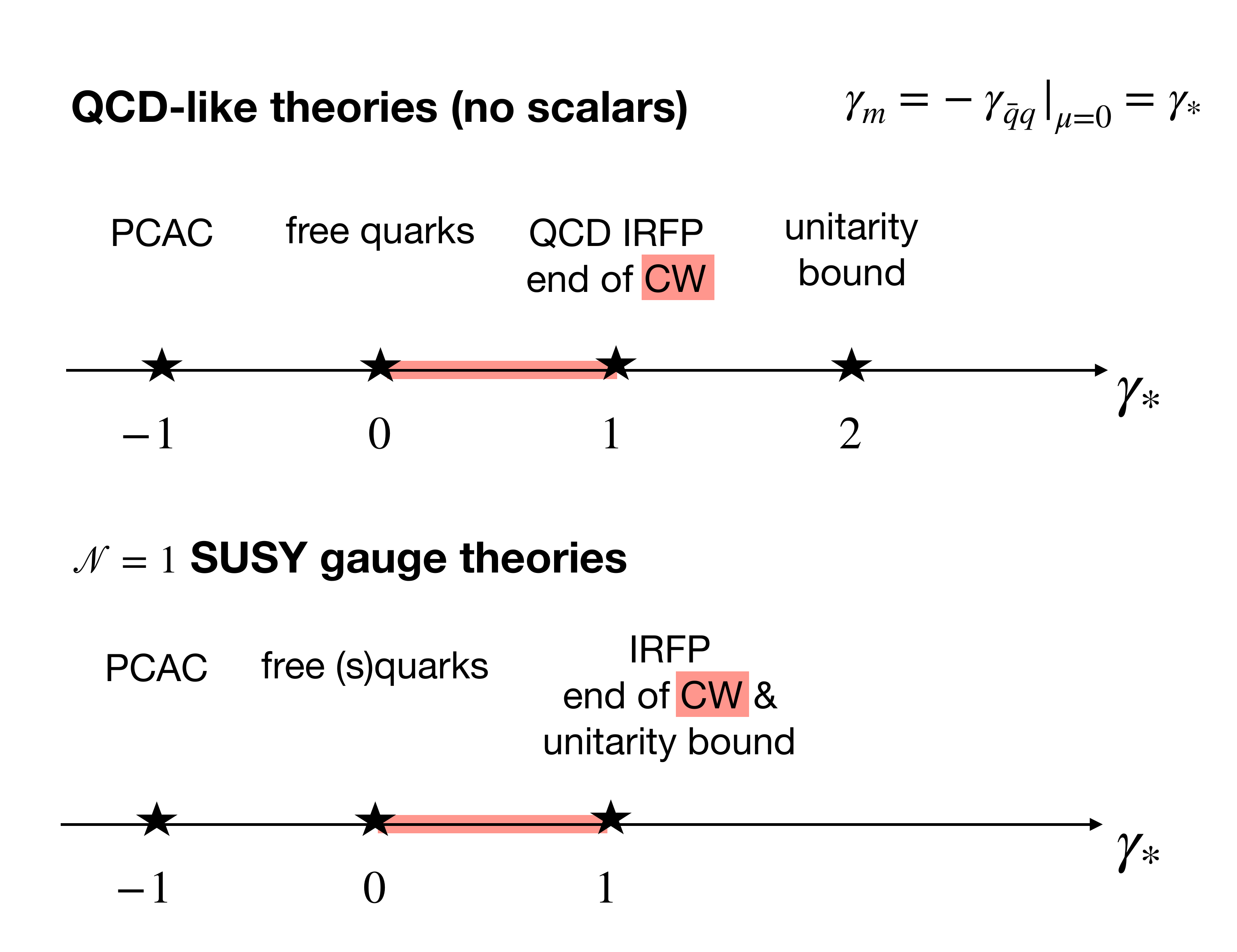}
\end{center}
\vspace{-0.5cm}
	\caption{\small 
	Range of possible IRFP anomalous dimension $\gast$. 
	 As emphasised in the main text, integer values seem to play a special role. 
	The value of $\gast$ is bounded from above by the unitarity bound 
	$\gast \leq 2(1)$, $\De_{\bar qq} = 3- \gast \geq 1$($\De_{\bar QQ } = 2 - \gast \geq 1$)  \cite{Mack:1975je} 
	in QCD-like theories (${\cal N} =1$ SUSY).  
	The lower bound $\gast > -1$ comes from the requirement of soft breaking 
	such that the  PCAC is not spoiled \cite{Wilson:1969zs}. 
	The value $\gast = 0$ corresponds to the trivial FP at the upper end of the conformal window cf. 
	\FIG\ref{fig:overview}. As the number of flavours is lowered,  $\gast $ raises and as it reaches  $\gast =1$, 
	chiral symmetry breaking sets in marking the  
	  lower end of the conformal window. This is true in ${\cal N} =1$, cf. footnote \ref{foot:SUSY}, 
	and in this paper  this is  conjectured to hold in QCD-like theories  as well. 
	The peculiarity of ${\cal N} =1$ is that the unitarity bound and the end of the conformal window coalesce whereas 
	this does not seem to be the case in QCD-like theories.} 
	\label{fig:gast}
\end{figure}

 Let us try to put this result into perspective, before rederiving it in two different ways.
 First it is noted that, $\gast =1$ is considerably below the unitarity bound $\gast \leq 2$, which follows from $\De_{\bar qq} = 3 - \gast  \geq 1$ \cite{Mack:1975je}.  
 The result gives rise  to the following picture.  
 For $\gast =0$ or $\De_{S^a} = 3$  it corresponds to 
 two free fermions, whereas for $\gast =1$ or  $\De_{S^a} = 2$ it describes  two free scalar  pions 
 and finally for $\gast = 2$ or $\De_{S^a} = 1$, when  reaching the unitarity bound, it is equivalent to  
 one free scalar particle \cite{Jost:1965yxu}.   The message seems to be that for integer powers of the scaling dimension, the theory lends itself to a  free particle interpretation, cf. \FIG\ref{fig:gast}.
 Note that the gauge theories only seem to make use of the  $[0,1]$-range in $\gast$, 
 which corresponds to only a third of the allowed range  $-1 \leq \gast \leq 2$ in the non-supersymmetric case. 
 
 Of course, \eqref{eq:chiPT} cannot be viewed as  novel from the $\chi$PT-viewpoint   as it 
is simply the LO-analysis.  However, what is new is the way in which this is realised 
in the gauge theory. 
The free pions are IR-dual to a gauge theory with a strongly coupled IRFP; strongly coupled since 
the anomalous dimension is large. 
These types of  interpretations  hold in many  EFT formulations of weakly coupled ultraviolet Lagrangians, 
and  may be regarded as the very purpose of the EFT-programme when the microscopic formulation is known.  
 
 This suggest the following picture 
 for the conformal window.  
 At the upper end  $\gast = 0$ and then $\gast$ increases 
 as $N_f$ is lowered  
 and when  $\gast =1 $ is reached chiral symmetry is broken and confinement sets in. 
 The anomalous dimension $\gast$, then remains one in the entire domain of the CD phase.
 As mentioned in the introduction the latter could be or not be identical to QCD itself.  
 The result \eqref{eq:main} is consistent  with  ${\cal N}=1$ supersymmetric gauge theories, as mentioned previously.

\subsection{Scaling of the pion mass implies  $\gast = 1$ ($\mq \neq 0$)}
\label{sec:gast}

In what follows the IRFP is perturbed by a non-vanishing quark mass $\mq$.
Even though the quark mass is scheme-dependent, the physics can be analysed by 
tracking powers of the rescaled bare mass.  
This is the standard method of hyperscaling  extensively applied to the conformal window \cite{DelDebbio:2010ze,DelDebbio:2010jy,Patella:2012da,DelDebbio:2013qta,Marcarelli:2022sbb} where hadrons appear when a  
quark mass term is introduced.\footnote{\label{foot:pic} 
The idea is that for $\mq \neq 0$ the quarks decouple leaving behind pure Yang-Mills which is known to confine \cite{Miransky:1998dh}. Hence there are hadrons and hadronic observables which however need to vanish when $\mq \to 0$. The way this happens is dictated by the RG \cite{DelDebbio:2010ze,DelDebbio:2010jy,DelDebbio:2013qta}. }  
The difference in the scenario at hand is that  chiral symmetry is spontaneously broken, 
and this introduces a natural cutoff scale $\La = 4 \pi F_\pi$ 
\cite{Manohar:1983md}. The quantity   $F_\pi \approx 93 \MeV$ in QCD is the pion decay 
constant and the order 
parameter of chiral symmetry breaking  \cite{Donoghue:1992dd,Leutwyler:1993iq,Scherer:2012xha}.   
We assume that   the $\chi$PT-cutoff    $\La$  
does not affect the LO $\mq$-behaviour of the pion mass, which is natural from the viewpoint of $\chi$PT itself which 
is organised in a $1/\La$-expansion. 
Under this assumption  the behaviour  of the pion mass is governed by  hyperscaling due to the RG, in the same way as in the conformal window. 
 The result, perhaps most cleanly derived in \cite{DelDebbio:2010jy}, is 
\begin{equation}
\label{eq:RG}
m^2_\pi|_{\text{RG}} \propto \mq^{\frac{2}{1+\gast}} \;,
\end{equation}
where $\gast$ is the previously introduced mass anomalous dimension at the FP. 
In QCD the linear behaviour 
\begin{equation}
\label{eq:QCD}
m^2_\pi|_{\text{QCD}} \propto \mq  \;,
\end{equation}
is deducible  in many ways such as from the  GMOR-relation \cite{Gell-Mann:1968hlm}, derived in  
 \APP\ref{app:softDouble}  from a double soft-pion theorem.  Since 
 \EQ\eqref{eq:QCD} holds in QCD, we  assume it would in a CD-QCD phase as well. 
 The dilaton is not affecting the  LO pion mass. 
Hence, equating \EQs \eqref{eq:RG} and \eqref{eq:QCD} implies the central result 
in $ \gast = 1$ \eqref{eq:main} once more. 

\subsection{Trace Anomaly  and Feynman-Hellmann theorem  $(\mq \neq 0)$}
\label{sec:FHTEMT}

The goal of this section is to show that the trace anomaly and the Feynman-Hellmann theorem are compatible 
if $\gast =1$ for an IRFP upon applying the formula 
\begin{equation}
\label{eq:start}
 2 m_\pi^2 =  \matel{\pi^a(p) }{\TEMT}{\pi^a(p)}  \;, \quad a \text{ fixed} \;.
\end{equation} 
The validity of \eqref{eq:start} when a dilaton is added will be commented on in \SEC\ref{sec:dilaton}.

\subsubsection{The $T^{\rho}_{\;\;\rho}$-anomaly and  renormalisation group invariant combinations}
\label{sec:RG}

The part relevant to physical matrix elements of the trace anomaly 
reads\footnote{The trace anomaly was first observed in correlation functions 
\cite{Crewther:1972kn,Chanowitz:1972vd,Chanowitz:1972da} and subsequently worked out in detail 
\cite{Minkowski:1976en,Adler:1976zt,Nielsen:1977sy,Collins:1976yq}  
including equation of motions  and BRST exact terms arising upon gauge fixing.}
 \begin{equation}
 \label{eq:TEMTphys}
 \TEMT|_{\text{phys}} =  \frac{\be}{2 g} G^2 + \sum_q \mq(1\pl \ga_m) 
    \bar{q}q  \;,
 \end{equation}
 where all quantities including the composite operators are renormalised. 
 
 An important aspect is that $m \bar qq$ is an RG invariant as mentioned previously. 
 Since $\TEMT$ is an RG invariant, the following two combinations 
 \begin{equation}
 \label{eq:inv1} 
 \Oone =  \frac{\be}{2g}G^2+ \sum_q  \ga_m  \mq \bar qq    \;, \quad   \Otwo = \sum_q  \mq \bar qq\;,
\end{equation}
 are RG invariants, or equally so, with $\de \ga  \equiv\ga_m- \gast   $
 \begin{equation}
 \label{eq:inv2}
 \Oone' =  \de \ga \, \sum_q  \mq \bar qq + \frac{\be}{2g}G^2   \;, \quad   \Otwo' =  (1+ \gast) \sum_q \mq \bar qq \; ,
\end{equation} 
 since the FP value $\gast$ is an RG invariant.   
Using \eqref{eq:start} to \eqref{eq:TEMTphys}, 
the $\ORD(\mq)$-contribution  then follows
\begin{equation}
\label{eq:TEMTm}
2 m_\pi^2  = (1+ \gast) \sum_q  \mq  \matel{{\pi}}{\bar qq}{{\pi}} + \ORD(\mq^2) \;,
\end{equation}  
The statement of \eqref{eq:TEMTm} is that $O_2'$ is the leading operator in the quark mass 
and that $O_1'$ is suppressed.  
Expanding  around the FP  in $\de g = (g-g^*)$, 
$\frac{d}{d\ln \mu} \de g =\be(g) = \best' \de g + \ORD(\de g^2)$, reveals that
\begin{equation}
\de g \propto \mu^{\best'} \propto \mq^{\frac{\best'}{1+\gast}} \;,
\end{equation}
where in the last equality  hyperscaling has been used,  which can been derived  from  blocking transformations  \cite{DelDebbio:2010jy}. 
If $\best' > 0$ then all the missing terms in \eqref{eq:TEMTm} are  power-suppressed in $\mq$. If $\best' =0$ 
then they are only logarithmically suppressed.\footnote{From the viewpoint of  $\chi$PT 
the relative corrections  are of order $\ORD(m_q \ln m_q)$, e.g. \cite{Scherer:2012xha,Leutwyler:1993iq}, and thus it is possible that the 
RG-analysis does not reveal the true next-to-LO behaviour. This is not relevant for 
the point we are making but worthwhile to investigate further.}   
The region $\best'<0$ is not allowed as otherwise there is no IRFP.
We therefore conclude that it is justified to neglect the $\de \ga$- and $\be$-terms to LO.

\subsubsection{The $T^{0}_{\;\;0} = H$ viewpoint -- Feynman-Hellmann theorem}
\label{sec:FH}

The Feynman-Hellmann theorem  \cite{Clusius1941EinfhrungID,Feynman:1939zza}   offers a way to obtain 
the LO quark mass dependence directly from the Hamiltonian $\de H_m =  \sum_q \mq  \bar q q$ by differentiation 
in $\mq$.  It is technically convenient to use  states, 
 $\langle \hat{\pi}(p') | \hat{\pi}(p) \rangle  =  (2\pi)^3  \de^{(3)}(\veci{p} -  \veci{p}')$, which are  normalised 
 in non-relativistic manner.  
One can  switch to the usual  states by $ \state{{\pi}} = \state{\hat{\pi}} \sqrt{2 E_\pi}$ after the $\mq$-differentiation. 
The Feynman-Hellmann formula  implies
\begin{equation}
 \partial_{\ln \mq} E_\pi  =
 \sum_q \mq \matel{\hat{\pi}}{\bar qq}{\hat{\pi}} + \ORD(\mq^2) \;,
\end{equation}
where $E_\pi = \matel{\hat{\pi}}{H}{\hat{\pi}}$, $V \leftrightarrow (2\pi)^3 \de^{(3)}(0)$ and $\partial_{\mq} \langle \hat{\pi}(p') | \hat{\pi}(p) \rangle =0$
have been used.  Switching back to standard pion states $\state{\pi}$ and using $\partial_{\mq}  E_\pi^2 =  \partial_{\mq} m_\pi^2$,  which follows from the $\mq$-independence of the  $3$-momentum $\vec{p}$, one obtains  
\begin{equation}
\label{eq:E}
 \partial_{\ln \mq} 2 m_\pi^2 = 2 \sum_q  \mq \matel{{\pi}}{\bar qq}{{\pi}} + \ORD(\mq^2) \;.
\end{equation} 
Further  assuming $m_\pi^2 = \ORD(\mq)$ then gives\footnote{The use of the Feynman-Hellmann theorem and its derivative is crucial. If one were to use the Hamiltonian, written schematically as 
$H = \vec{E}^2+\vec{B}^2 + \sum_q \mq \bar qq$, then applying the states would result in 
$2 E_\pi = \matel{\pi}{\vec{E}^2+\vec{B}^2}{\pi} + \matel{\pi}{\sum_q \mq \bar qq}{\pi}$, where the momentum dependence 
of $E_\pi$ has to reside in the electromagnetic $\vec{E}^2+\vec{B}^2$ matrix element, cf. \cite{Prochazka:2013vya} for related discussions. }
\begin{equation}
\label{eq:simple}
2 m_\pi^2   = 2 \sum_q  \mq  \matel{{\pi}}{\bar qq}{{\pi}} + \ORD(\mq^2) \;,
\end{equation}
the formula linear in the quark mass. 
 The formula \eqref{eq:simple}  in itself is 
not new and has been used and derived in the literature frequently 
e.g. \cite{Gasser:1982ap,DelDebbio:2013sta}.
The correctness of \eqref{eq:simple} is verified in \APP\ref{app:softDouble} by reproducing 
the GMOR-relation \cite{Gell-Mann:1968hlm}. This is important as an incorrect  numerical prefactor, 
by matching to the trace anomaly below, would give an incorrect $\gast$.

\subsubsection{Matching the two mass formulae}
\label{sec:together}

The two mass formulae, \eqref{eq:TEMTm} and  \eqref{eq:simple}, 
 are, once more, compatible with each other if and only if $\gast = 1$.   
We consider this an important result since the  assumption is weaker than in  \SEC\ref{sec:gast}.\footnote{In the setting of the conformal window, with quark mass deformation (\cite{DelDebbio:2013qta} and footnote \ref{foot:pic}), both approaches lead to 
$m^2_\phi \propto (\mq)^{2/(1+\gast)}$ where $\phi$ stands for any hadron.  The situation is though 
different in the case at hand because of the cutoff scale mentioned above.}
In that section we assumed that the renormalisation behaviour (hyperscaling), 
in the pion sector at  LO in the quark mass, is unaffected by the presence of the $\chi$PT cutoff $\La = 4 \pi F_\pi$. 
Here we merely assumed that the $\be$- and $\de \ga$-terms can be neglected in the vicinity 
of the FP.  As the RG-scale $\mu$ can be made arbitrarily small  seems a lesser  assumption and 
thus more satisfying in our view.

 It seems worthwhile to point out that 
independent of whether there is an IRFP or not,  $\matel{\pi}{O_1}{\pi} = \matel{\pi}{O_2}{\pi} + \ORD(\mq^2) $.  
Or in more familiar notation
\begin{equation}
\label{eq:must}
\matel{\pi}{\frac{\be}{2g}G^2  +  \sum_q  \mq\ga_m \bar qq }{\pi}  =   \matel{\pi}{\sum_q \mq \bar qq }{\pi} + \ORD(\mq^2) \;,
\end{equation}
assures that the trace anomaly and Feynman-Hellmann derivation of the LO pion mass 
are consistent with each other.
The solution for an IRFP $\be \to \best =0$ and $\ga_m \to \gast =1 $ 
is a straightforward one.
Other solutions, not related to an IRFP, demand a specific interplay between the $\be$- and $\ga$-term.  This is 
though perfectly possible since $O_1$ in \eqref{eq:inv1} is an RG invariant.

\section{Brief Comments on the Addition of a Dilaton}
\label{sec:dilaton}

The results obtained did not make use of the presence of a dilaton. Conversely, if pion physics can be interpreted by an IRFP then this suggests that 
a dilaton could be present and it is a valid question whether the latter would impact on any of the results obtained. 
Let us refer for practical reasons to the dilaton as the lightest state in the $J^{PC} =0^{++}$ flavour singlet channel. 
If the dilaton remains massive in the limit where the explicit symmetry breaking is removed, $\mq \to 0$, then it can simply be integrated out 
in the deep-IR and everything remains the same.  If on the other hand it becomes massless in that limit then a closer inspection is needed. 
We proceed case by case.
\begin{itemize}
\item For the long-distance correlator in \SEC\ref{sec:deep} there would be a diagram where one 
of the propagating pions in \FIG\ref{fig:dias} is replaced by a dilaton. The gives raise to the 
same $1/x^4$ behaviour  as in \eqref{eq:chiPT} but with a different prefactor, in particular  $\Nf$-independent, which excludes exact cancellation. 
Hence the conclusions remain unchanged. 
\item The hyperscaling argument of \SEC\ref{sec:gast} is also unaltered but it implies  in turn $m_D^2 \propto \mq$ in the same way as it does for the pion.
\item  The matching of the trace anomaly and the Feynman-Hellmann theorem in \SEC\ref{sec:FHTEMT} is the more subtle and requires some care.  In the case of a massless dilaton the standard formula  $2 m_\phi^2 =  \matel{\phi}{\TEMT}{\phi}$, where $\phi$ is a physical state, 
cannot be used because of the dilaton pole  \cite{DelDebbio:2021xwu}.  However, in the case of the pion \eqref{eq:start}  holds
since the effect of the dilaton pole for massless states, such as the pion,  is undone by its  coupling 
to pions. Concretely, 
\begin{equation}
\label{eq:ZTEMT}
 \matel{\pi^a(p')}{T_{\mu\nu} }{\pi^a(p)} \supset
c\,  (q_\mu q_\nu - q^2 \eta_{\mu\nu})  \frac{g_{D\pi\pi}}{q^2- m_D^2}     \;, \quad q \equiv p - p'
\end{equation}
where $c = \text{const} \times F_D$ and $g_{D\pi\pi}$ is given by  \cite{Ellis:1970yd,Crewther:2015dpa,Zwicky:2023fay,RZdilaton}
\begin{equation}
g_{D\pi\pi} =  \frac{1}{F_D}( 2 q^2 + 2(1-\gast) m_\pi^2 + \ORD(\mq^2)) \;,
\end{equation}
where $F_D$ is the dilaton decay constant as defined in  \cite{DelDebbio:2021xwu}. The $q^2$-dependence originates from the pion kinetic term to which the dilaton couples.  
Taking the trace in \eqref{eq:ZTEMT} and  the limit $q^2 \to 0$ we learn that this term does not contribute. 
The pion and the dilaton are both massive due to $\mq \neq 0$, which 
would be enforced in a systematic EFT approach. 
The behaviour for   massive hadrons is qualitatively different since $g_{D\phi\phi} \propto m_\phi^2/F_D = \ORD(\La)$, here for a scalar $\phi$, 
does not vanish any limit.  It is precisely this behaviour that gives rise to the vanishing of the trace of the EMT for a 
massless dilaton \cite{DelDebbio:2021xwu}.

A further point of concern is that the dilaton could alter the evaluation of the matrix element  
$\matel{\pi^a}{\bar qq}{\pi^a}$ in \APP\ref{app:softDouble}. The dilaton contribution to this matrix element is analogous to 
\eqref{eq:ZTEMT} is $\propto  g_{D\pi\pi}/(q^2- m_D^2)$ (without the $q$-dependent prefactor).
Assuming $\gast \to 1$ and $q^2 \to 0$  we learn that this term does not contribute. 
Two remarks are in order. First to use $\gast =1$ is legitimate since it has already been 
concluded by equating  \eqref{eq:TEMTm} and  \eqref{eq:simple}. Second it is crucial to keep $m_D \neq 0$, due to $\mq \neq 0$. 
\end{itemize}
We infer from our considerations that the addition of a (massless) dilaton does not alter the results.

 \section{Summary and Conclusions}
 \label{sec:conc}
 
In this work we have offered an interpretation of  low energy pion physics in terms of a strongly coupled infrared fixed point of QCD-like gauge theory. 
Colloquially speaking, this means that  the  infrared states such as the pions experience the world as a conformal field theory in the deep infrared.  
Comparing observables in the conformal or renormalisation group picture with standard pion physics 
we deduced  in three  ways that the quark-mass anomalous dimension  takes on the value $\gast =1$, at the fixed point. 
Namely, 
 \begin{itemize}
 \item by requiring consistency between the leading order $\chi$PT and the CFT-interpretation of the
 adjoint scalar correlator $\vev{\OSa(x) \OSa(0) }$ in \SEC\ref{sec:deep}.
  \item by renormalisation-group arguments and assuming that the $\chi$PT cutoff $\La = 4 \pi F_\pi$  does not affect  
 leading quark-mass behaviour of the pion in \SEC\ref{sec:gast}.
 \item by requiring consistency between the  trace anomaly and the Feynman-Hellmann theorem  
in \SEC\ref{sec:FHTEMT}.\footnote{Note that \EQ\eqref{eq:must}  must hold in the chirally broken phase irrespective 
of the fixed point interpretation.}
 \end{itemize}
These arguments are largely independent and thus any of the three could have served 
as a starting point for the paper.  Perhaps, the third point is the strongest as it only relies on the near fixed point behaviour.   
The important point is, though, that by assuming an infrared fixed point  we were able to derive 
 internally consistent results. This is no substitute for a proof. 
 In fact there are at least the three possibilities shown in \FIG\ref{fig:overview}.  The scenario is not realised in any gauge theory (left), it is realised in some area 
 outside the conformal window  (centre) or it is  identical to the standard QCD-type theories (right). 
 
As a byproduct this led us to  conjecture that $\gast =1$ marks the end of the conformal window. 
This last point is consistent with lattice Monte Carlo computations \cite{DelDebbio:2010zz,Nogradi:2016qek,Witzel:2019jbe},  in particular the dilaton-EFT 
fits in \cite{Appelquist:2017wcg,Appelquist:2019lgk}, 
perturbative computations  \cite{Ryttov:2016asb,Ryttov:2017kmx}, gap equations \cite{Yamawaki:1985zg,Appelquist:1988yc,Cohen:1988sq}, 
walking technicolor phenomenology \cite{Hill:2002ap,Sannino:2009za,Appelquist:2003hn}, holographic approaches \cite{Jarvinen:2011qe,Alvares:2012kr} and ${\cal N}=1$ supersymmetry \cite{Intriligator:1995au,Shifman:1999mk,Terning:2006bq}. 
  However, these works do not interpret  the pion physics below the boundary by an infrared fixed point, which is the main point 
of our work.  From a certain perspective our work is more closely related to the pre-QCD work 
 \cite{Isham:1970xz,Isham:1970gz,
Ellis:1970yd,Ellis:1971sa,Crewther:1970gx,Crewther:1971bt}  or its revival a decade ago \cite{Crewther:2013vea,Crewther:2015dpa}. The difference to these papers is that there is a definite statement 
about the scaling of the most important operators.

Another way to look at the proposal  is to notice that QCD in the deep infrared  is described by the  free field theory of pions  and is thus scale invariant.\footnote{If the dilaton is added 
 then the theory becomes at least classically conformal \cite{Zwicky:2023fay}, circumventing the Goldstone improvement problem.} 
 This makes the fixed point interpretation look natural,  and is indeed assumed   in the context of the $a$-theorem, e.g. \cite{Cardy:1988cwa}. 
 The $\chi$PT gauge-theory matching can be seen as infrared duality of weak and strong coupling theories, c.f. 
 \SEC\ref{sec:match}, which are often 
 the motivation for an effective-field-theory programme. That these types of dualities are more fundamental, 
 might be related to the Seiberg-dualities \cite{Intriligator:1995au,Shifman:1999mk,Terning:2006bq}, which in turn gave new motivation 
 to the fascinating idea of hidden local symmetry, e.g.  \cite{Yamawaki:2018jvy,Komargodski:2010mc,Abel:2012un}.  
 
 There are other factors supporting the  infrared-fixed-point picture. 
The Goldstone improvement  \cite{Zwicky:2023fay}, 
 ameliorates the convergence of the integrals of the  $a$-theorem   
 for QCD-like theories  \cite{Komargodski:2011vj,Luty:2012ww}.\footnote{A sufficiently fast vanishing $\TEMT$ in the IR is a necessary condition for the integral formulae to converge.} Or dense nuclear interactions support a fixed point scenario  \cite{Ma:2019ery,Rho:2021zwm,Brown:1991kk}.
The addition of the dilaton sector to be discussed in \cite{RZdilaton} will offer other ways to test the scenario. 
 As mentioned earlier the dilaton candidate in QCD is
the broad $f_0(500)$ meson.  Importantly, if the Higgs sector is to be replaced by a gauge theory then its dilaton can take on the role of a Higgs which is  hard to distinguish 
 from the Standard Model one.   
 This has been appreciated  since a long time within the gauge theory setting  
 e.g. \cite{Cata:2018wzl} and without concrete ultraviolet completion e.g. 
 \cite{Goldberger:2007zk,Bellazzini:2012vz}  Our work strengthens this case 
 considerably  and   identifies  in $S = \bar qq$ the presumably most relevant operator 
 as its scaling dimension assumes  $\De_{\bar qq} =2$ as a consequence of $\gast =1$. 
 A further advantage of gauge theories for a dilaton sector is that they can be 
explored with analytic tools and lattice Monte Carlo simulations serving as 
a  laboratory to further ideas in a concrete setting.

\acknowledgments 
RZ is supported by a CERN associateship and an STFC Consolidated Grant, ST/P0000630/1.   
I am grateful to  Steve Abel, Tim Cohen, Gilberto Colangelo, Matthew McCullough, 
Poul Damgaard, Luigi Del Debbio 
John Donoghue, John Ellis, Nick Evans, Max Hansen, Shoji Hashimoto, Andreas J\"uttner, Daniel Litim, Tom  Melia,  Kohtaroh Miura, Agostino Patella, J\'er\'emie Quevillon, Mannque Rho, Francesco Sannino, Misha Shifman, Christopher Smith, 
Lewis Tunstall  and Koichi Yamawaki  for correspondence or discussions. 

\appendix

\section{Conventions}
\label{app:conv}

The Minkowski metric $\mink_{\mu\nu}$  reads  $\textrm{diag}(1,-1,-1,-1)$.
The Lagrangian of the gauge theory  is given by 
\begin{equation}
\label{eq:LQCD}
{\cal L} = - \frac{1}{4}G^2  + \sum_q \mq \bar q (i \slashed{D} -\mq) q \;,
\end{equation}
where $G^2 = G_{\mu\nu}^A G^{A \mu\nu}$ is the field strength tensor and $A$  the adjoint index  
of the gauge group.  The $N_f$ quark flavours are assumed to be degenerate in mass.
The beta function is defined by ${\be} = \frac{d}{d\ln \mu}   g $ and the mass anomalous dimension 
is given by $\ga_m  = - \frac{d}{d\ln \mu} \ln \mq $. Quantities at the FP  are designated by a star e.g. 
$\gast = \ga_m|_{\mu=0}$ \eqref{eq:gast}. QED is omitted 
even though the massless photon is definitely an IR degree of freedom but it does not change the picture considerably 
as it is weakly coupled in the IR. The $SU(N)$ flavour symmetry generators $T^a$ are normalised as 
$T^a T^b = \frac{1}{2 N_c}  \de^{ab} \mathbb{1}_{N_c} + \frac{1}{2} d^{abc} T^c + \frac{i}{2} f^{abc} T^c$, 
$ \Tr[T^a T^b] = \frac{1}{2}\delta^{ab}$ and $f/d^{abc}$ are the totally anti/symmetric tensors.

\section{Conformal versus Scale Invariance}
\label{app:conscale}

Scale and conformal invariance are not distinguished  in this work as it is widely believed that the former implies the latter for theories like QCD (and most non-exotic $d=4$ theories) cf. Ref.~\cite{Nakayama:2013is}  for a review. 
A scale invariant theory is one where 
$\TEMT= \partial \cdot V$ such that $J^{\dil}_\mu = x^\nu T_{\mu\nu}- V_\mu$ is conserved. 
Since the scaling dimension of the trace of the EMT is $d$, the one of the virial current has to be $d-1$ which is highly 
non-generic as it, usually, requires the protection of a symmetry.
  
 \section{Soft-pion Theorem} 
  \label{app:soft}
  
 Since the soft-pion theorem is important in the main text, we reproduce 
 its form from the textbook \cite{Donoghue:1992dd}
 \begin{equation}
 \label{eq:soft}
\matel{\pi^a(q) \be}{{\cal O}(0)}{\al} = - \frac{i}{F_\pi} \matel{ \be}{[Q_5^a,{\cal O}(0)]}{\al}  + \lim_{q \to 0} i q \cdot R^a \;,
 \end{equation}
where the square brackets denote the commutator. 
Above $\al$ and $\be$ are other physical states and $R^a$ is the so-called remainder 
 \begin{equation}
 \label{eq:R}
 R^a_\mu =    - \frac{i}{F_\pi}\int d^dx \, e^{i q\cdot x} \matel{ \be}{T J_{5\mu}^a(x) {\cal O}(0)}{\al}  \;,
 \end{equation}
 which  vanishes unless there are  intermediate states degenerate with either $\al$ or $\be$.\footnote{We have checked that it vanishes in the cases at hand and will therefore 
 not discuss it any further.  A case where the remainder is relevant is 
 the matrix element $\matel{N^a \pi^b(q)}{J_\mu^c}{ N^{*d} }$. The $\state{N^{a'}}$  is degenerate 
 and one can use the Callan-Treiman relation, due to the chiral Ward identity,  to infer
   $\lim_{q \to 0} q^\mu \matel{ N^a}{ J_{5\mu}^b}{N^d} \neq 0$, implying the non-vanishing of the remainder.}
\EQ\eqref{eq:soft} is straightforward to derive from correlation functions using a dispersive representation.

\subsection{The GMOR-relation  from  double soft-pion theorem} 
\label{app:softDouble}
 
 In \SEC\ref{sec:together} it was concluded that the trace anomaly and the Feynman-Hellmann theorem 
 imply $\gast =1$ but this relies in particular that the \emph{prefactor} in \EQ\eqref{eq:simple} is correct. 
 This can be verified by making the link to the celebrated GMOR-relation \cite{Gell-Mann:1968hlm} of QCD.  
 The procedure is to apply the soft theorem, summarised above, twice to eliminate the pions.
Applying it once  results in 
\begin{equation}
\label{eq:mpi2}
   m_\pi^2 = \sum_{q} \mq  \matel{\pi^a}{\bar qq}{\pi^a} =  \frac{-\mq}{F_\pi} 
  \matel{0}{ i [ Q_5^a , \bar q \,\mathbb{1}_{N_f} q]}{\pi^a}  
 = \frac{ 2 m_q }{F_\pi} \matel{0}{P^a}{\pi^a}  \;,
\end{equation}
where $P^a = \bar q  i\ga_5 T^a q$ as previously, and $\sum_q \bar qq \to \bar q \, \mathbb{1}_{N_f} q$ as it is a more suitable notation to evaluate the commutator.  The remainder \eqref{eq:R} can be omitted since it is zero. This is not obvious when a dilaton is present as commented on  in \SEC\ref{sec:dilaton}.
Applying the soft theorem to \eqref{eq:mpi2} once more, using 
$d^{abc} \vev{\bar q T^c q} =0$,  one gets
\begin{equation}
\matel{0}{P^b}{\pi^a} =  - \frac{1}{F_\pi} \matel{0}{i [Q_5^a,P^b]}{0} = - \frac{1}{F_\pi}  \vev{\bar qq}\de^{ab} \;,
\end{equation}
which combines into  
 \begin{equation}
m_\pi^2 F_\pi^2 =   - 2m_q \vev{\bar qq} \;,
\end{equation} 
the  GMOR-relation \cite{Gell-Mann:1968hlm,Donoghue:1992dd,Leutwyler:1993iq,Scherer:2012xha}.
This completes the task of this appendix.

\bibliographystyle{utphys}
\bibliography{../Dil-refs.bib}

\end{document}